\documentclass[preprint,12pt]{elsarticle}

\usepackage{amssymb}
\usepackage{mathtools}
\usepackage{amsmath}

\journal{Optics and Laser Technology}

\begin{document}

\begin{frontmatter}



\title{Quasi-calibration method for structured light system with auxiliary camera}


\author[inst1]{Seung-Jae Son}
\author[inst2]{Yatong An\corref{cor}}
\author[inst1]{Jae-Sang Hyun\corref{cor}}
\cortext[cor]{Corresponding Author}

\affiliation[inst1]{organization={Department of Mechanical Engineering, Yonsei University},
            postcode={03722}, 
            state={Seoul},
            country={South Korea}}

\affiliation[inst2]{organization={Meta Reality Labs},
            city={Redmond},
            postcode={98052}, 
            state={Washington},
            country={USA}}

\begin{abstract}
The structured light projection technique is a representative active method for 3-D reconstruction, but many researchers face challenges with the intricate projector calibration process. To address this complexity, we employs an additional camera, temporarily referred to as the \emph{auxiliary camera,} to eliminate the need for projector calibration. The auxiliary camera aids in constructing rational model equations, enabling the generation of world coordinates based on absolute phase information. Once calibration is complete, the auxiliary camera can be removed, mitigating occlusion issues and allowing the system to maintain its compact single-camera, single-projector design. Our approach not only resolves the common problem of calibrating projectors in digital fringe projection systems but also enhances the feasibility of diverse-shaped 3D imaging systems that utilize fringe projection, all without the need for the complex projector calibration process.

\end{abstract}


\begin{highlights}
\item The proposed method introduces a novel approach to structured light system calibration, significantly enhancing efficiency without compromising on accuracy.

\item The proposed method incorporates an auxiliary camera to facilitate calibration, accommodating both digital and non-digital illumination sources effectively.

\item The proposed method offers a reduction in the time and resources required for structured light system calibration, making advanced applications more accessible and cost-efficient.

\item The proposed method has been validated through extensive experimental analyses.
\end{highlights}

\end{frontmatter}



\section{Introduction}
Fringe projection profilometry (FPP) technique typically includes a digital projector as an active illuminator to generate fringe signals and a digital camera to receive the reflected signals\cite{schreiber2000theory,scharstein2003high,du2007three,ishiyama2007absolute, gorthi2010fringe,brauer2011using,li2013multiview}. Since the digital projector allows per-pixel control of the projection image and the 3-D point cloud can be retrieved by deterministic one-to-one mapping problems, the digital fringe projection (DFP) system is widely used for inspecting the product(s) in manufacturing or scanning highly complex object in medicine and forensic science required to be high-quality 3-D point cloud\cite{pan2009phase, lei2009flexible, zhang2010recent, zhang2012review, suresh2018high}.

Many commonly used hand-held 3-D scanners and highly accurate 3-D scanners with static setups adopt two cameras and one projector system\cite{weise2007fast,wang2012robust,jang2013structured,tao2016real,eiriksson2016precision}. The main principle of this system is the same as the standard stereo vision, which is a representative passive 3-D reconstruction method. The processor finds the correspondence between two images and calculates 3-D points with the triangulation method. The only difference is that the fringe images from the projector give the processor the information of the correspondence and make it possible to calculate the 3-D point cloud pixel by pixel. However, if the system tries to capture the object(s) having large depth gaps, it is difficult to reconstruct 3-D, and it results in occlusion since all three optical devices, two cameras, and one projector, should look at the point to be reconstructed. To overcome the limitation, Zhang developed the projector calibration method by making the projector an inverse camera and simplified the configuration of the DFP system\cite{zhang2006novel}. Compared to camera calibration, which has been deeply researched in the computer vision field, projector calibration requires elaborate procedures since it can only see through the camera's field of view. By doing so, the pinhole model for the projector and camera can be uniquely solved, which means that the system can generate the 3-D geometry pixel by pixel with only two optical devices. Even if it is successful to retrieve the 3-D geometry, there is the distortion problem at the corner of the 3-D geometry when the projector imaging plane reprojects the images to the camera's imaging plane. To resolve the issue, many researchers capture the flat plane as the reference plane and make a look-up table to compensate for the error caused by that distortion\cite{li2008accurate,luo2014simple,feng2014high}. However, there is still a problem in that the accuracy of the DFP system is determined depending on the fabrication accuracy of the flat plane and it is difficult to guarantee which part of the 3-D geometry of the flat plane is correct. Recently, Zhang developed a method to minimize the error by defining the relationship between the world coordinate and the phase value extracted from the fringe images\cite{zhang2021flexible}. By using the fact that the 3-D geometry of the object can be calculated by the phase information of the captured fringe images, they set the relationship between the phase value and the world coordinates with the third-order polynomial equation. The measurement error of the system converges, and the distortion effect at the corner becomes much less than before. Nevertheless, the tedious projector calibration procedure should be done before applying this method, meaning that a special projector that does not follow the pinhole model cannot be used, such as the mechanical projector developed by Hyun for high-speed and sub-pixel accuracy\cite{hyun2018high}.
To address the aforementioned limitations, in this research, we propose a novel calibration method for fringe projection systems. Initially, one auxiliary camera is temporarily added to the system. By implementing an auxiliary camera, the system can reconstruct 3-D geometry using the conventional stereo vision method with the projector\cite{zhang2000flexible}. Based on the calculated 3-D geometry, we can define the relationship between the fringe information and the world coordinate with respect to the main camera by using the approximate rational model equation\cite{vargas2023pixel}. Then, there is no need to keep the auxiliary camera, and without the projector calibration, the system can reconstruct 3-D pixel by pixel. This approach effectively alleviates the problems of camera distortion and skew. Furthermore, our calibration method is not limited to calibrating the DFP system. Instead, it is possible to calibrate any type of FPP system that generates an absolute phase map. Experimental outcomes confirm that our method delivers a 3-D reconstruction resolution almost identical to traditional techniques, all while simplifying the calibration process.

The contents of this paper is as follows. In Section~\ref{sec2}, we introduce the theoretical background of our research. Section~\ref{sec3} describes the detailed procedure of our experiment. In addition, Section~\ref{sec3} also includes a rigid evaluation of our research to verify our results. Section~\ref{sec4} summarizes this work.

\section{\label{sec2}Principle}
\subsection{N-step phase shifting method}
The FPP adopts the concept of phase to find corresponding pixels between the camera and the projector. Using phase instead of intensity has many advantages in reconstructing 3-D geometry; the reconstructed 3-D result becomes more noise tolerant because the system designer can manage the light sources. Moreover, it is less affected by surface reflection and color of the texture.
To find the phase value via the system, a set of fringe pattern should be projected onto the target and captured by the camera. The projected pattern should include at least three frames of phase-shifted fringe patterns. This methodology, which utilizes multiple frames to recover phase values, is called the N-step phase shifting method. It finds the phase value $\phi$ by solving the equations as follows,
\begin{equation}
\begin{split}
 I_1(u,v) &= I'(u,v)+I''(u,v)\cos(\phi(u,v)+\delta_1),\\ 
 I_2(u,v) &= I'(u,v)+I''(u,v)\cos(\phi(u,v)+\delta_2),\\ 
 I_3(u,v) &= I'(u,v)+I''(u,v)\cos(\phi(u,v)+\delta_3),\\
 & \vdotswithin{=}\\
 I_k(u,v) &= I'(u,v) + I''(u,v)\cos(\phi(u,v)+\delta_k)
\end{split}
\label{eq:1}
\end{equation}
where $I'(u,v)$ is the average intensity and $I''(u,v)$ is the intensity modulation of corresponding pixel $(u,v)$ at the $k-$th frame. For example, if there are \(N\) frames of phase-shifted images, phase \(\phi\) can be found by solving Eq.~(\ref{eq:1}) for $\phi$ with the least square method. Then the solution \(\phi(u,v)\) is represented as follows,
\begin{equation}
\begin{split}
\phi(u,v) =-\tan^{-1}(\frac{\sum_{k=1}^{N}I_k\sin\delta_k}{\sum_{k=1}^{N}I_k\cos\delta_k})
\end{split}
\end{equation}
However, due to the characteristic of the inverse tangent function, the retrieved phase ranges from \(-\pi\) to \(\pi\) with \(2\pi\) discontinuities. 
The discontinuities result in having multiple points with the same phase value, meaning that it cannot set one-to-one correspondence between the camera and projector pixel. A procedure called phase unwrapping is required to resolve the issue, making the retrieved phase continuous and absolute. The phase values are unwrapped with the equation as follows,
\begin{equation}
\Phi(u,v) =\phi(u,v)+2\pi\times K(u,v)
\end{equation}
where $\Phi(u,v)$ represents the unwrapped phase, the phase without discontinuities, and $K(u,v)$ represents the proper integer added to the phase value to connect the discontinuities. Here, we have denoted the unwrapped phase as $\Phi(u,v)$ to distinguish it from the phase $\phi(u,v)$, which has discontinuities in each period. Temporal phase unwrapping, used in the following experiments, connects the discontinuities by projecting additional binary patterns containing the order $K(u,v)$ of the fringe period.

\subsection{Fringe projection profilometry system}
\begin{figure}[h!]
\centering\includegraphics[width=0.6\textwidth]{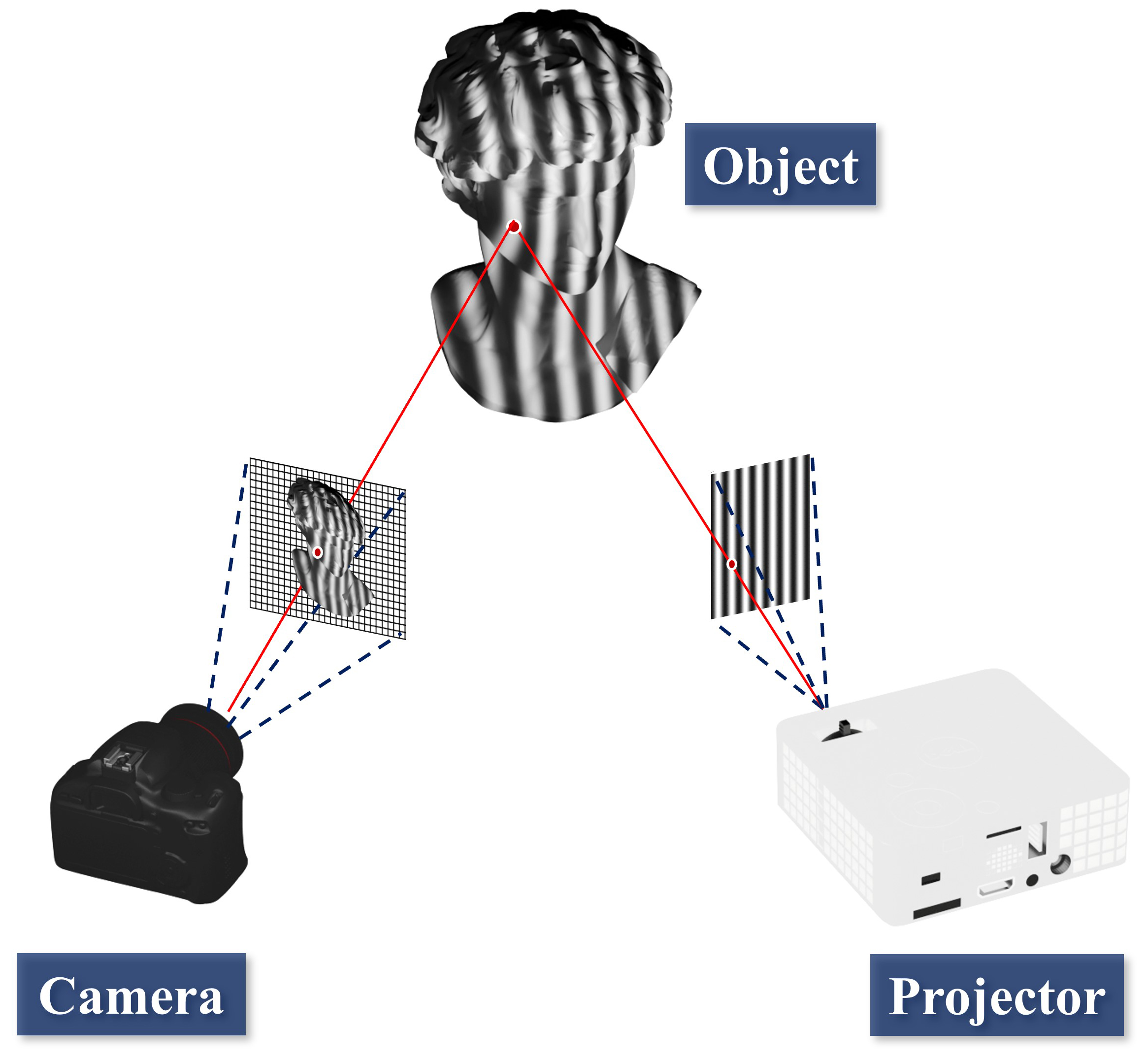}
\caption{Schematic of the fringe projection profilometry system}
\label{fig:systemSchematic}
\end{figure}

Every optical device is represented as a pinhole model. The pinhole model is a mathematical expression of the relationship between the 3-D world coordinates $(x,y,z)$ and the image pixel $(u,v)$. The linear pinhole model is represented mathematically as follows,
\begin{equation}
    s
    \begin{bmatrix}
    u \\ v \\ 1
    \end{bmatrix}
    = \textbf{A}
    \begin{bmatrix}
    \textbf{R} | \textbf{t}
    \end{bmatrix}
    \begin{bmatrix}
    x_{\text{w}} \\ y_{\text{w}} \\ z_{\text{w}} \\ 1
    \end{bmatrix},
\end{equation}
where $s$ is a scaling factor and $\textbf{A}$,$ \textbf{R}$ and $\textbf{t}$ are

\begin{equation}
 \textbf{A}= \begin{bmatrix}f_u & \gamma & c_u \\ 0 & f_v & c_v \\ 0 & 0 & 1 \end{bmatrix},\textbf{R} = \begin{bmatrix} r_{11} & r_{12} & r_{13} \\ r_{21} & r_{22} &r_{23} \\ r_{31} & r_{32} & r_{33} \end{bmatrix},\textbf{t} = \begin{bmatrix} t_1 \\ t_2 \\ t_3\end{bmatrix},
\end{equation}

respectively. $\textbf{A}$ represents the intrinsic parameters of the imaging device, which includes the effective focal length $f_u$ and $f_v$, principal point $c_u$, and $c_v$ and skew factor $\gamma$. For the research level camera, $\gamma=0$.
If we define projection matrix $\textbf{P}$ as
\begin{equation}
    \textbf{P} = \textbf{A}\begin{bmatrix} \textbf{R}|\textbf{t} \end{bmatrix} = \begin{bmatrix} p_{11} & p_{12} & p_{13} & p_{14} \\ p_{21} & p_{22} & p_{23} & p_{24} \\ p_{31} & p_{32} & p_{33} & p_{34} \end{bmatrix},
\end{equation}
linear pinhole model can be turned into,
\begin{equation}
    s\begin{bmatrix} u \\ v \\ 1 \end{bmatrix} = \textbf{P}\begin{bmatrix} x_{\text{w}} \\ y_{\text{w}} \\ z_{\text{w}} \\ 1 \end{bmatrix}.
\end{equation}
Then, the linear pinhole model of the camera and projector can be expressed as follows,
\begin{equation}
    s^{\text{c}}\begin{bmatrix} u^{\text{c}} & v^{\text{c}} & 1 \end{bmatrix}^{\text{T}} = \textbf{P}^{\text{c}}\begin{bmatrix} x^{\text{w}} & y^{\text{w}} & z^{\text{w}} & 1 \end{bmatrix}^{\text{T}},
\label{eq:8}
\end{equation}
\begin{equation}
    s^\text{p}\begin{bmatrix} u^{\text{p}} & v^{\text{p}} & 1 \end{bmatrix}^{\text{T}} = \textbf{P}^{\text{p}}\begin{bmatrix} x^{\text{w}} & y^{\text{w}} & z^{\text{w}} & 1 \end{bmatrix}^{\text{T}}.
\label{eq:9}
\end{equation}
where the superscripts ${\text{c}}$ and ${\text{p}}$ represent the camera and projector respectively, and the superscript $\text{T}$ represents the transpose of a matrix. Fig.~\ref{fig:systemSchematic} shows the common configuration of the FPP system. If the FPP system is calibrated, the projection matrix $\textbf{P}^{\text{c}}$ and $\textbf{P}^{\text{p}}$ is determined. Then, the FPP system finds the world coordinates by solving Eq.~(\ref{eq:8}) and Eq.~(\ref{eq:9}). If the camera pixel $u=u_0$ and $v=v_0$ is taken into account when solving the equation, the corresponding world coordinates are described as follows,
\begin{equation}
    \begin{bmatrix} x_0 \\ y_0 \\ z_0 \end{bmatrix} = \begin{bmatrix} p^{\text{c}}_{31}u_0-p^{\text{c}}_{11} & p^{\text{c}}_{32}u_0-p^{\text{c}}_{12} & p^{\text{c}}_{33}u_0-p^{\text{c}}_{13} \\ p^{\text{c}}_{31}v_0-p^{\text{c}}_{21} & p^{\text{c}}_{32}v_0-p^{\text{c}}_{22} & p^{\text{c}}_{33}v_0-p^{\text{c}}_{23} \\ p^{\text{p}}_{31}u^{\text{p}}-p^{\text{p}}_{11} & p^{\text{p}}_{32}u^{\text{p}}-p^{\text{p}}_{12} & p^{\text{p}}_{33}u^{\text{p}}-p^{\text{p}}_{13} \end{bmatrix}^{-1} \begin{bmatrix} p^{\text{c}}_{14}-p^{\text{c}}_{34}u_0 \\ p^{\text{c}}_{24}-p^{\text{c}}_{34}v_0 \\ p^{\text{p}}_{14}-p^{\text{p}}_{34}u^{\text{p}} \end{bmatrix}
\label{eq:10}
\end{equation}
where $p^{\text{c}}$ and $p^{\text{p}}$ represent elements of the projection matrix of the camera and projector. If the fringes are projected, phase values will be generated with the N-step phase-shifting method. When the vertical fringes are projected, the phase values on the camera pixel can define the corresponding projector pixel as follows,

\begin{equation}
    u_{\text{p}} = \Phi \times T / (2\pi)
\label{eq:11}
\end{equation}
where $T$ represents the fringe periods in pixels. If we replace $u_{\text{p}}$ with Eq.~(\ref{eq:11}), and solve Eq.~(\ref{eq:10}) with respect to $(x_0,y_0,z_0)$, the world coordinates can be expressed as follows,
\begin{equation}
    \begin{bmatrix} x_0 \\ y_0 \\ z_0 \end{bmatrix} = \begin{bmatrix} x_{0}(\Phi) \\ y_{0}(\Phi) \\ z_{0}(\Phi) \end{bmatrix}.
\end{equation}
which show that the world coordinates$(x_0,y_0,z_0)$ are a function of the absolute phase value $\Phi$ because the other values are constant.

\subsection{Pixel-wise calibration method}
Previous studies in fringe projection profilometry have shown that the phase value \(\Phi\) and the world coordinates $(x,y,z)$ have a certain relationship. According to Zhang\cite{zhang2021flexible}, the relationship can be approximated by the third-order polynomials for each pixel. The polynomial model defines the relationship between the phase value $\Phi$ and world coordinates $(x_{\text{w}},y_{\text{w}},z_{\text{w}})$ in each camera pixel as follows
\begin{equation}
    x^{\text{c}}(\Phi)=a_0+a_1\Phi+a_2\Phi^2+a_3\Phi^3
\end{equation}
\begin{equation}
    y^{\text{c}}(\Phi)=b_0+b_1\Phi+b_2\Phi^2+b_3\Phi^3
\end{equation}
\begin{equation}
    z^{\text{c}}(\Phi)=c_0+c_1\Phi+c_2\Phi^2+c_3\Phi^3
\end{equation}
where $a_0, ..., a_3, b_0, ..., b_3, c_0, ..., c_3$, are the coefficients for the polynomial model, in which each pixel has its own unique value. A fringe projection profilometry system calibrated with a pixel-wise polynomial model has been shown to reconstruct the 3-D geometry with higher accuracy, especially on the edges of the image. However, the polynomial model performs well only in the short depth range. To overcome that limitation, Vargas et al. calibrated the fringe projection profilometry system using another pixel-wise calibration model called the rational model\cite{vargas2023pixel}. The relationship between $\Phi$ and $(x^{\text{c}},y^{\text{c}},z^{\text{c}})$ in the rational model is as follows,
\begin{equation}
    z^{\text{c}}=\frac{m_{0}+m_{1}\Phi}{1+m_{2}\Phi}
\end{equation}
\begin{equation}
    x^{\text{c}}=m_{3}z^{c}
\end{equation}
\begin{equation}
    y^{\text{c}}=m_{4}z^{c}
\end{equation}

where $m_0, m_1, m_2, m_3, m_4$, are the calibration coefficients for each pixel. While a polynomial model has been assumed based on the experimental data without backbone equations, the rational model was derived from the pin-hole model equation. Compared with the polynomial model, the rational model has achieved a larger depth range of the reconstruction area while maintaining the high accuracy of the pixel-wise calibration method.
\subsection{Proposed method}
\begin{figure}[h!]
\centering\includegraphics[width=0.9\textwidth]{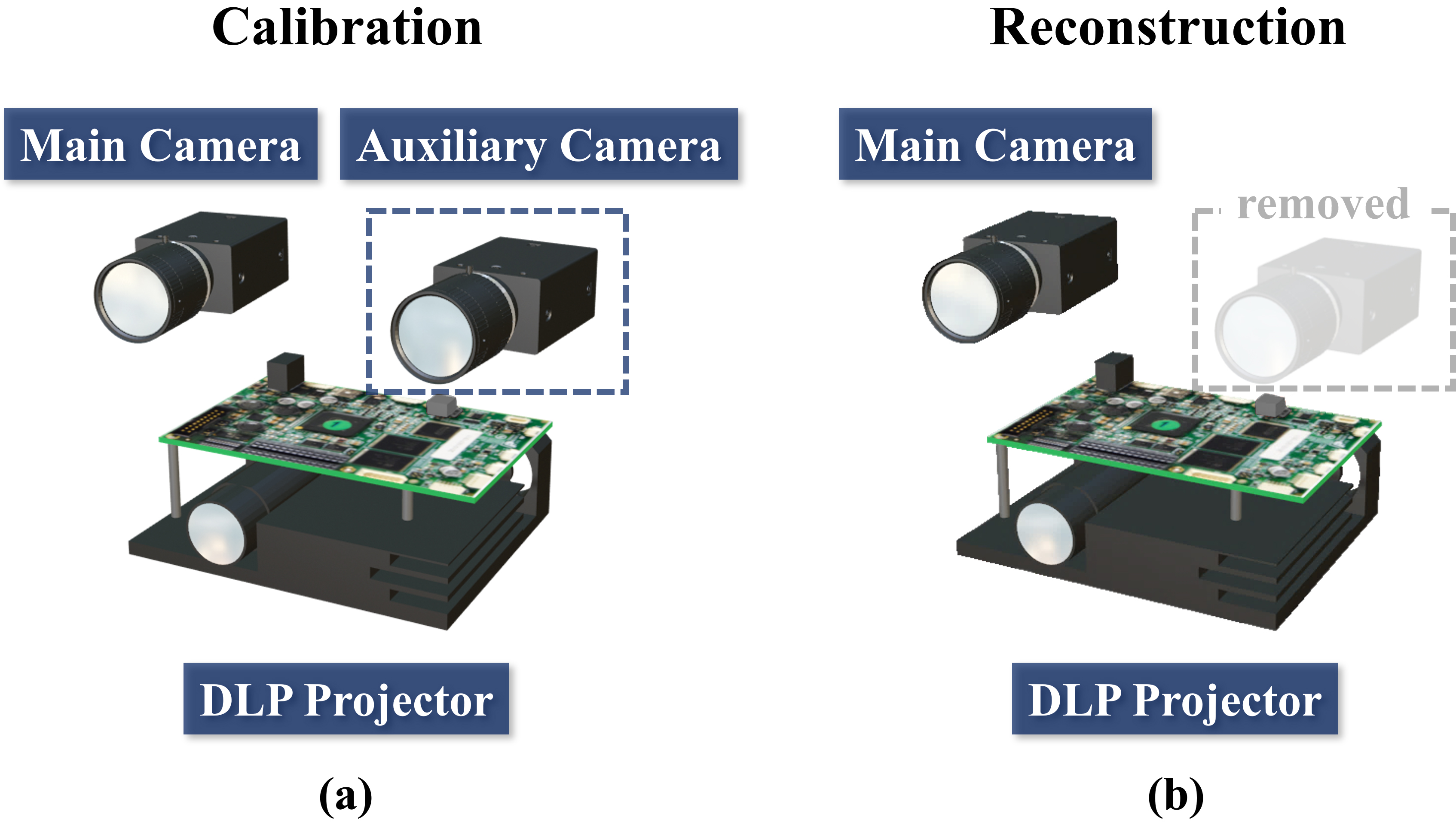}
\caption{System configuration of the proposed method, consists of two cameras and one DLP projector. After the calibration, only the main camera and the DLP projector are used for reconstructing the geometry.}
\label{fig:calibrationReconstruction}
\end{figure}
Calibration of a projector in a FPP system is more complicated than calibration of a camera because the projector is output device. Despite previous efforts, such as Zhang's projector calibration method utilizing camera-captured images\cite{zhang2006novel}, calibrating the projector still remains a sophisticated process\cite{wang2013optimal,li2014adaptive,zheng2017phase, li2017high, feng2021calibration}. In this context, we have developed a system calibration method for FPP system that does not require a projector calibration process throughout the sequence. Moreover, the preceding FPP systems were constrained to employing expensive Digital Light Processing (DLP) projectors as illumination devices. DLP projectors are required because two-directional fringe projection is required for calibration, even though the reconstruction requires only one-directional fringe pattern. However, the proposed method overcomes this limitation by enhancing the system's adaptability depending on the equipment conditions. Calibration of the projector-camera system will be possible, only if the projection can generate a one-directional phase map with repeatability.
This method uses one projector and two cameras. Since we can remove the second camera after the calibration process, we named the second camera \emph{an auxiliary camera}, and the first one, \emph{a main camera}. Fig.~\ref{fig:calibrationReconstruction} compares the system configuration during the calibration and reconstruction. The proposed method calibrates the system by finding the relationship between a phase value and world coordinates in each pixel of the main camera. Here, to attain world coordinates in each pixel, the auxiliary camera is temporarily used during the calibration. If the system is calibrated, 3-D geometry can be measured without it. The main camera and projector acquire depth information using previously derived phase-coordinate relations. We named this novel method \emph{quasi-calibration method}. The proposed calibration process consists of the following steps.

\begin{itemize}
    \item \textit{Step 1:} Calibrate the main and \emph{auxiliary camera} with the standard stereo vision calibration method to obtain the intrinsic and extrinsic parameters.
    \item \textit{Step 2:} Find the 3-D geometry of the calibration board with stereo cameras based on a conventional stereo-matching algorithm with the phase information calculated by the N-step phase shifting algorithm.
    \item \textit{Step 3:} Determine the ideal planar model that fits the three-dimensional geometry of the calibration board by utilizing the plane fitting equation.
    \item \textit{Step 4:} Find the relationship between the 3-D geometry, $(x^{\text{c}},y^{\text{c}},z^{\text{c}})$, which is centered by the main camera and $\Phi$ for each pixel in the pixel-wise calibration model.
    \item \textit{Step 5:} Using only the main camera and digital projector, capture the fringe images of the object(s) to be measured and reconstruct the 3-D geometry based on the previously established relationship.
\end{itemize}

While finding the 3-D geometry of the calibration board (Step 2), single-directional fringe patterns are projected, and an absolute phase map is generated to give guidance on finding the corresponding pixels of each camera. However, the generated absolute phase map contains phase errors along the boundaries of the calibration circles, resulting from the high contrast between the white circle and a black background. To remove certain artifacts, the phase map was fitted on a smooth polynomial surface\cite{zhang2021flexible}. Additionally, before fitting the observation data to the pixel-wise calibration model, the closest flat plane was generated via the equation of the plane as following
\begin{equation}
    Ax+By+Cz+D = 0
\end{equation}
where the variables A, B, C, and D are determined to minimize the offset error between the observation and the ideal plane (Step 3). The coefficients of the pixel-wise calibration model are retrieved with the point cloud values of an ideal plane and absolute phase values (Step 4). Using an ideal plane instead of an initial observation clears out the inherent distortions of the optical devices. If coefficients of the pixel-wise calibration model are found, the system is calibrated at this moment, and the auxiliary camera can be removed if necessary. The 3-D shape of the target can be measured only with the projector and the main camera. The following experiment presents 3-D results of the FPP system calibrated with the proposed method. It is proved that this method calibrates the FPP system without projector calibration and successfully reconstructs a complex object. Furthermore, the following experiments conducted with a plastic slit illuminator show that the proposed method allows users to utilize wider types of fringe illumination devices.

\section{\label{sec3}Experiment}
 \begin{figure}[h!]
 \centering\includegraphics[width = 0.7\textwidth]{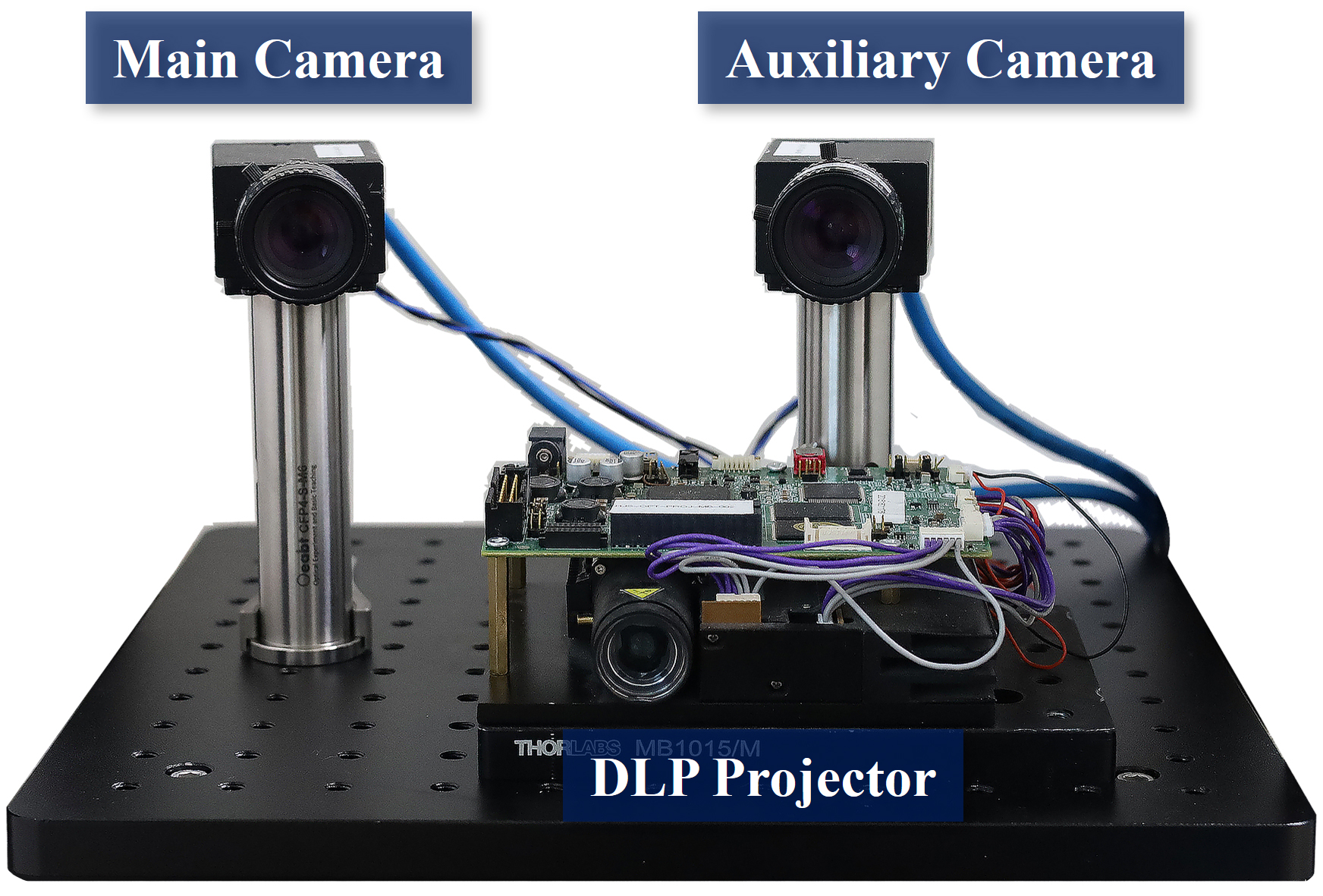}
 \caption{Experimental setup with DLP projector}
 \label{fig:system_setup}
 \end{figure}

 \begin{figure}[!t]
 \centering\includegraphics[width = 0.9\textwidth]{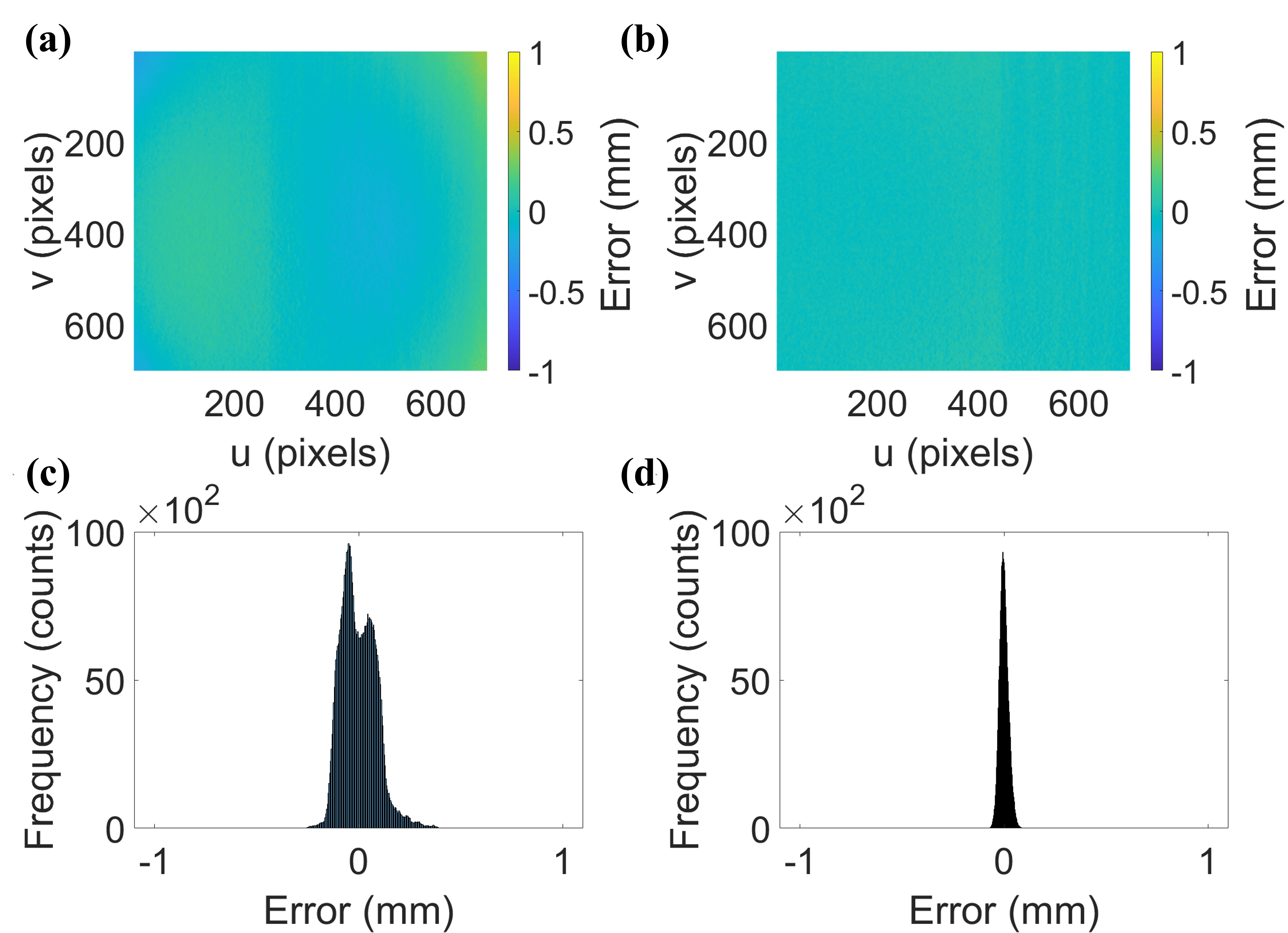}
 \caption{3-D reconstruction results of the flat plane. (a) Offset error map from the ideal plane to the measurement taken by the system calibrated with the conventional method. (b) Offset error map from the ideal plane to the measurement taken by the system calibrated with the proposed method. (c) Histogram of the offset error map of the conventional method. (d) Histogram of the offset error map of the proposed method.}
 \label{fig:mirrorOffset}
 \end{figure}

 \begin{figure}[!t]
 \centering\includegraphics[width = 0.9\textwidth]{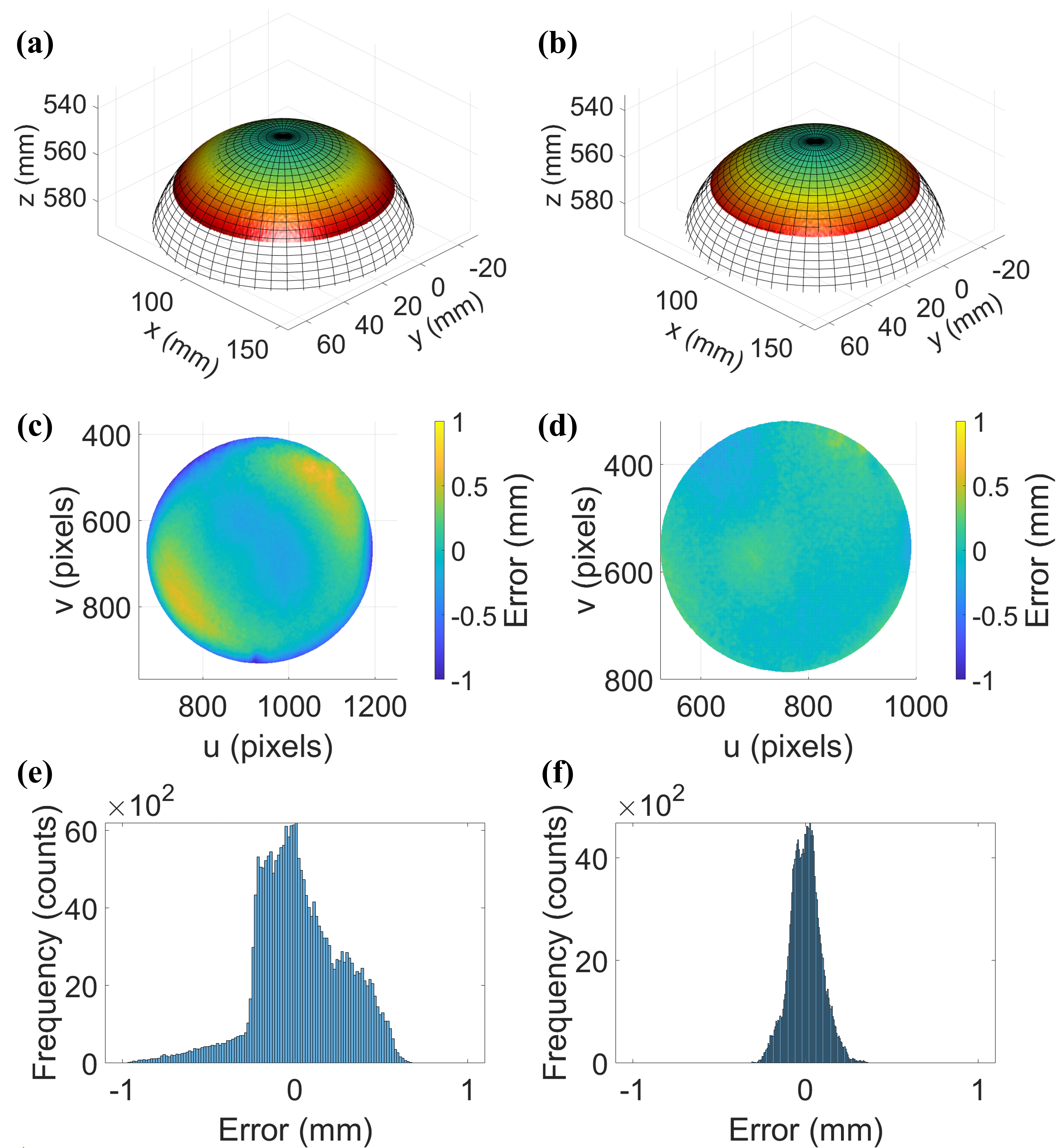}
 \caption{3-D reconstruction results of the sphere with a known-radius 50 mm. (a) 3-D reconstruction of the sphere taken with the system calibrated by the conventional method. (b) 3-D reconstruction of the sphere taken with the system calibrated by the proposed method. (c) Error map from the ideal sphere to the measurement taken by the system calibrated with the conventional method. (d) Error map from the ideal sphere to the measurement taken by the system calibrated with the proposed method. (e) Histogram of the error map of the conventional method. (f) Histogram of the error map of the proposed method.}
 \label{fig:sphereOffset}
 \end{figure}

 \begin{figure}[!htb]
 \centering\includegraphics[width = 0.9\textwidth]{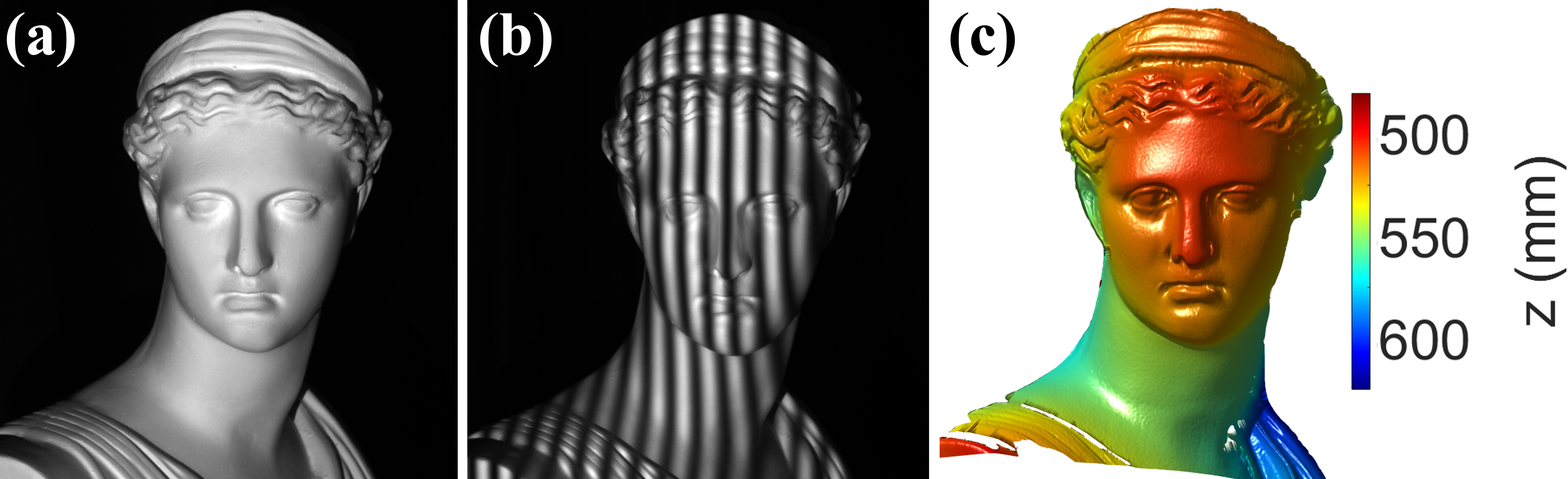}
 \caption{3-D reconstruction results of the statue. (a) Texture image of the statue. (b) Fringe projected image of the statue. (c) 3-D reconstruction of the statue taken with the system calibrated with the proposed method.}
 \label{fig:statueDLP}
 \end{figure}
 
 To validate the proposed method, we set up the system as follows: two CCD cameras (Model: FLIR GS3-U3-50S5M-C) with 12 mm focal length lens (Model: Computar M1214-MP2) and one digital light processing (DLP) projector (Model: Texas Instrument DLP LightCrafter 4500) as shown in Fig.~\ref{fig:system_setup}. The resolution of the camera was set to $1600\times1200$, and the projector's resolution was $912\times1140$.

 The first step of the proposed method is to calibrate the main camera and the auxiliary camera with the standard stereo vision method. Two cameras have taken 60 poses of the calibration board to calibrate the main camera and the auxiliary camera. Next, the 3-D geometry of the calibration board was reconstructed by employing the stereo camera and a projector.
 
 The projector projected vertical fringe patterns on the calibration board. By capturing those projected fringe patterns with two cameras, each camera generated a phase map based on the phase-shifting algorithm. The disparity between the two cameras was found based on the phase value. The projected patterns consisted of a set of 18 phase-shifted images to generate phase values with a fringe pitch of 18 pixels and 7 gray code images for unwrapping these phase values. Next, 20 poses of the calibration board were taken. The generated phase map was once fitted with the equation of the polynomial surface to remove artifacts resulting from large intensity deviations between the adjacent pixels around the edges of the calibration circles. Here, we used a fifth-order polynomial surface equation. Assuming that the calibration board is completely flat, the 3-D geometry of the plane was fitted to the equation of the plane. The relationship between the $(x^{\text{c}},y^{\text{c}},z^{\text{c}})$ values and the $\Phi$ values of each pixel were found with the rational model using the fitted data. Once the relationship has been found, the geometry was estimated with the main camera and projector. In other words, the auxiliary camera was no longer required to estimate the geometry. During geometry estimation, as in the previous pattern, 25 patterns of 18 phase-shifted images and 7 gray code images were projected onto the geometry.

 In order to verify the reconstruction accuracy of the proposed method, three objects, including the flat plane, sphere, and statue, were reconstructed with the main camera and projector. For comparison, the same objects were reconstructed using the system calibrated with the conventional FPP calibration method. Calibration images were attained via the same calibration board at the exact position for both calibration methods. Both methods were calibrated using the 20 poses of the calibration board over a range of $300\,\text{mm}<z^{\text{c}}<600\,\text{mm}$. However, the conventional calibration method required both horizontal and vertical patterns, whereas the proposed method required one directional fringe pattern. Therefore, additional horizontal patterns were projected after vertical ones onto each pose to calibrate the system by the conventional method. A total of 52 frames of calibration boards with projections for each pose were taken with the system. The 52 patterns include 25 vertical and 25 horizontal fringe patterns as well as additional black and white patterns. It is noteworthy that the following experimental data were filtered with 5$\times$5 Gaussian filter in the phase map domain to remove significant noise from the measurement. Fig.~\ref{fig:mirrorOffset} shows the flat plane reconstruction by each calibration method. Fig.~\ref{fig:mirrorOffset}(a) and Fig.~\ref{fig:mirrorOffset}(b) show the corresponding error map, which presents the offset from the measurement to the ideal plane. Fig.~\ref{fig:mirrorOffset}(c) and Fig.~\ref{fig:mirrorOffset}(d) show the histogram of the corresponding offset error map. The root-mean-square(RMS) error of the error map has been estimated to be 0.0860 mm for the conventional method and 0.0207 mm for the proposed method. The RMS error of the proposed method has the same level of accuracy as the latest calibration methods \cite{wang2022high,zhang2023flexible}. Fig.~\ref{fig:sphereOffset} shows the sphere reconstruction by each calibration method. Fig.~\ref{fig:sphereOffset}(a) and Fig.~\ref{fig:sphereOffset}(b) show the ideal sphere and estimated result, and Fig.~\ref{fig:sphereOffset}(c) and Fig.~\ref{fig:sphereOffset}(d) show the corresponding error map, which presents the offset from the measurement to the ideal sphere. Fig.~\ref{fig:sphereOffset}(e) and Fig.~\ref{fig:sphereOffset}(f) show the histogram of the corresponding offset error map. The estimated radius of the sphere was 50.05 mm with the proposed method, which is an error of 0.1\% from the ground truth radius of 50 mm. The estimated RMS error of the sphere was 0.1877 mm for the conventional method and 0.0901 mm for the proposed method. Fig.~\ref{fig:statueDLP} shows the results of reconstructing complex geometry, in this case, the statue of Artemis, using the proposed calibration method.

 \subsection{FPP system with plastic slit illuminator}

 \begin{figure}[!htb]
 \centering\includegraphics[width = \textwidth]{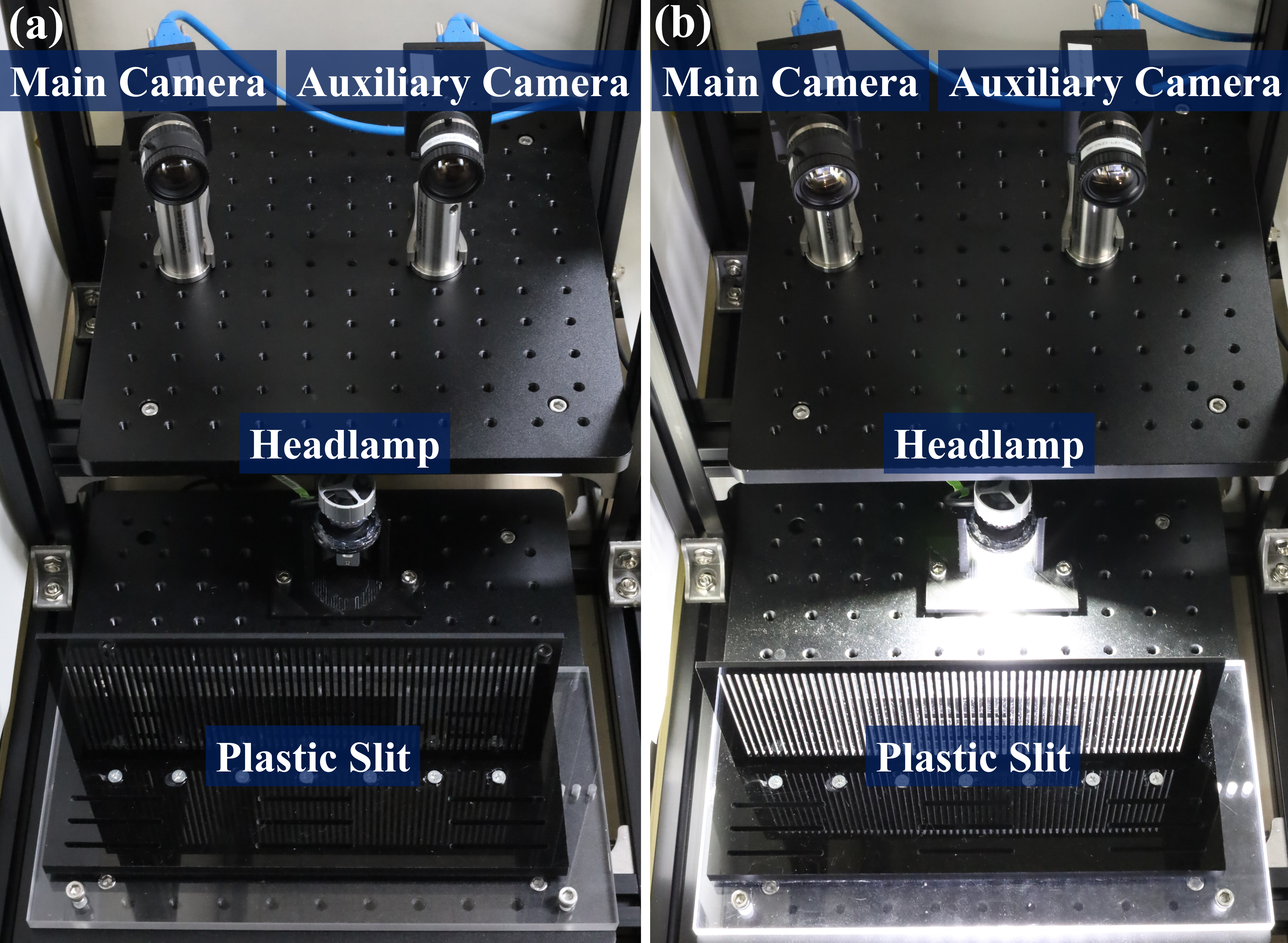}
 \caption{Experimental setup with the plastic slit illuminator. (a) Experimental setup with headlamp turned off. (b) Experimental setup with headlamp turned on.}
 \label{fig:plasticSetup}
 \end{figure}

 \begin{figure}[h!]
 \centering\includegraphics[width = \textwidth]{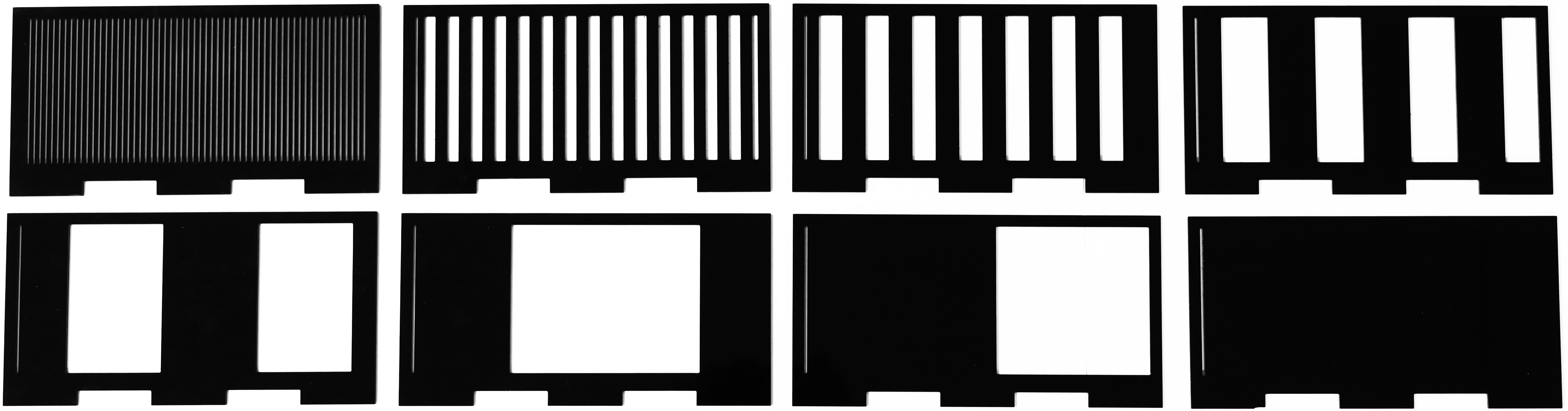}
 \caption{Plastic slits used in the experiment. The upper left slit was used for phase-shifting, and the period of the pattern was 4 mm. The remaining slits were used to generate a gray code pattern.}
 \label{fig:plasticSlit}
 \end{figure}
 
 \begin{figure}[h!]
 \centering\includegraphics[width = \textwidth]{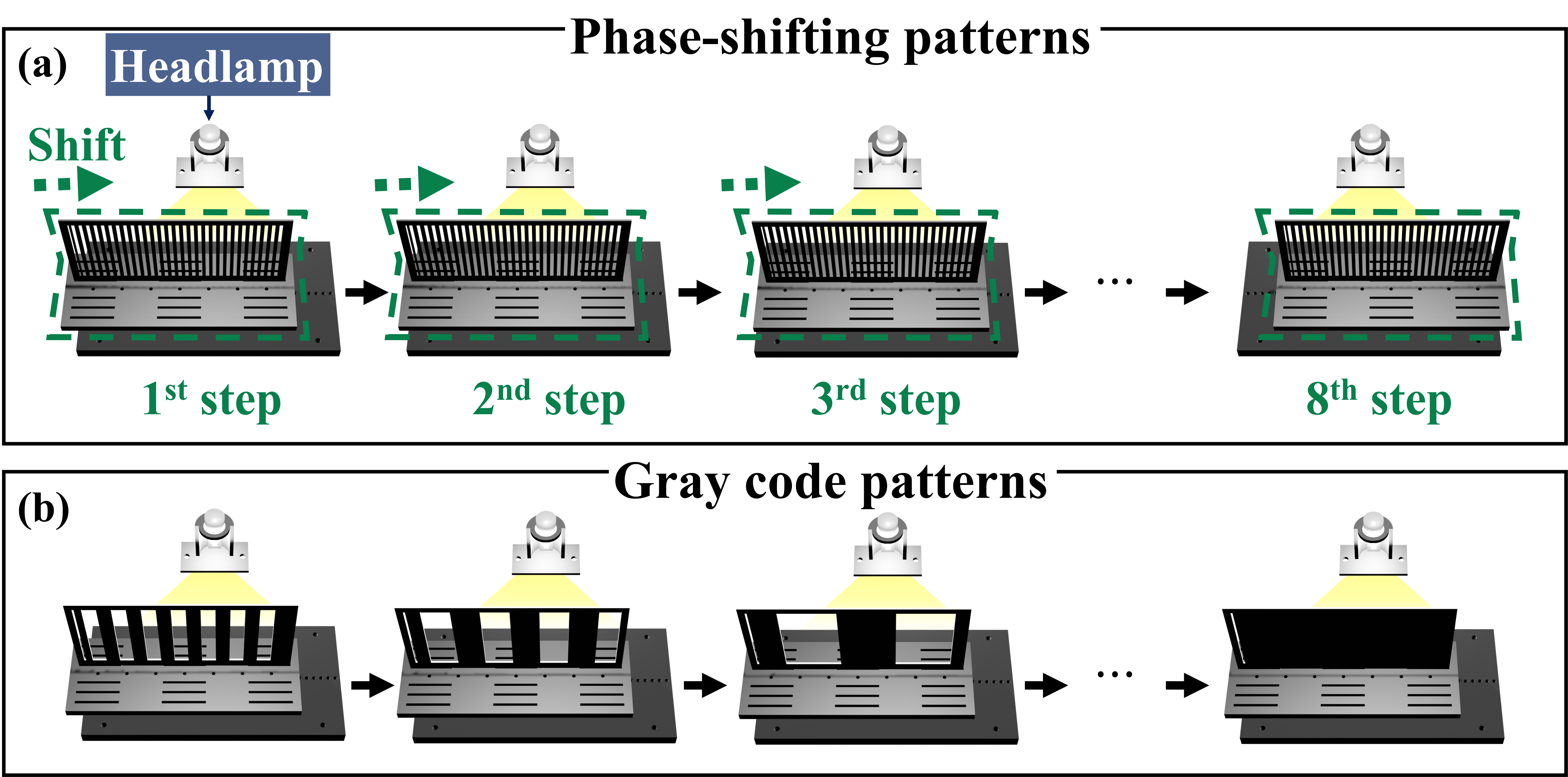}
 \caption{Illustration explaining how we manually moved plastic slit. (a) Phase-shifting patterns were projected by moving the plastic slit in a lateral direction. (b) Gray code patterns were projected by changing the plastic slits at the exact position.}
 \label{fig:plasticPatterns}
 \end{figure}
 
 \begin{figure}[!htb]
 \centering\includegraphics[width = \textwidth]{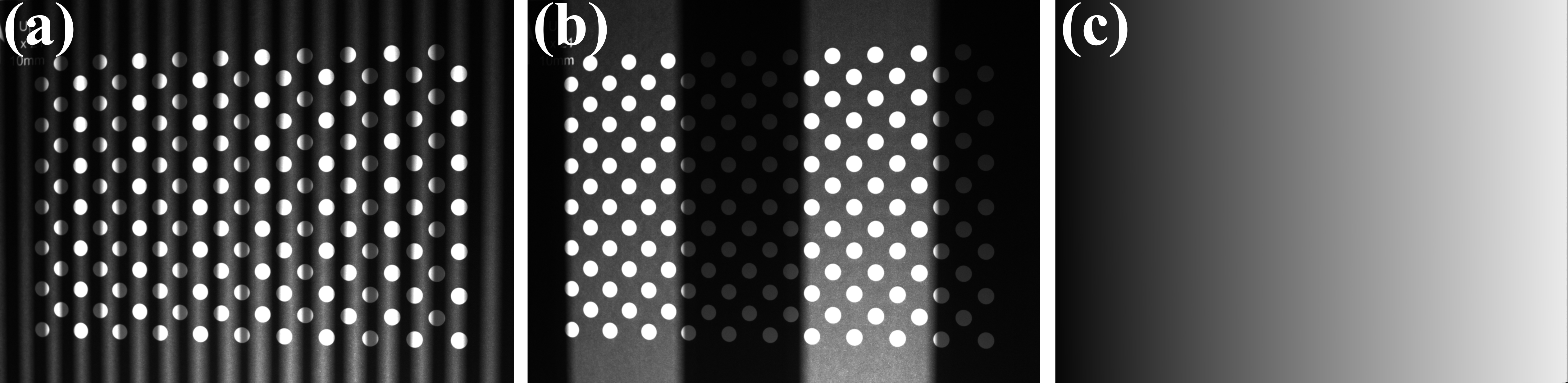}
 \caption{Images used to calibrate the plastic slit illumination system and corresponding phase map. Patterns were projected using the plastic slit illuminator. (a) Calibration board with N-step phase shifting pattern projected on it. (b) Calibration board with gray code pattern projected on it. (c) Phase map generated by N-step phase shifting method.}
 \label{fig:plasticFringes}
 \end{figure}

 \begin{figure}[!htb]
 \centering\includegraphics[width = 0.7\textwidth]{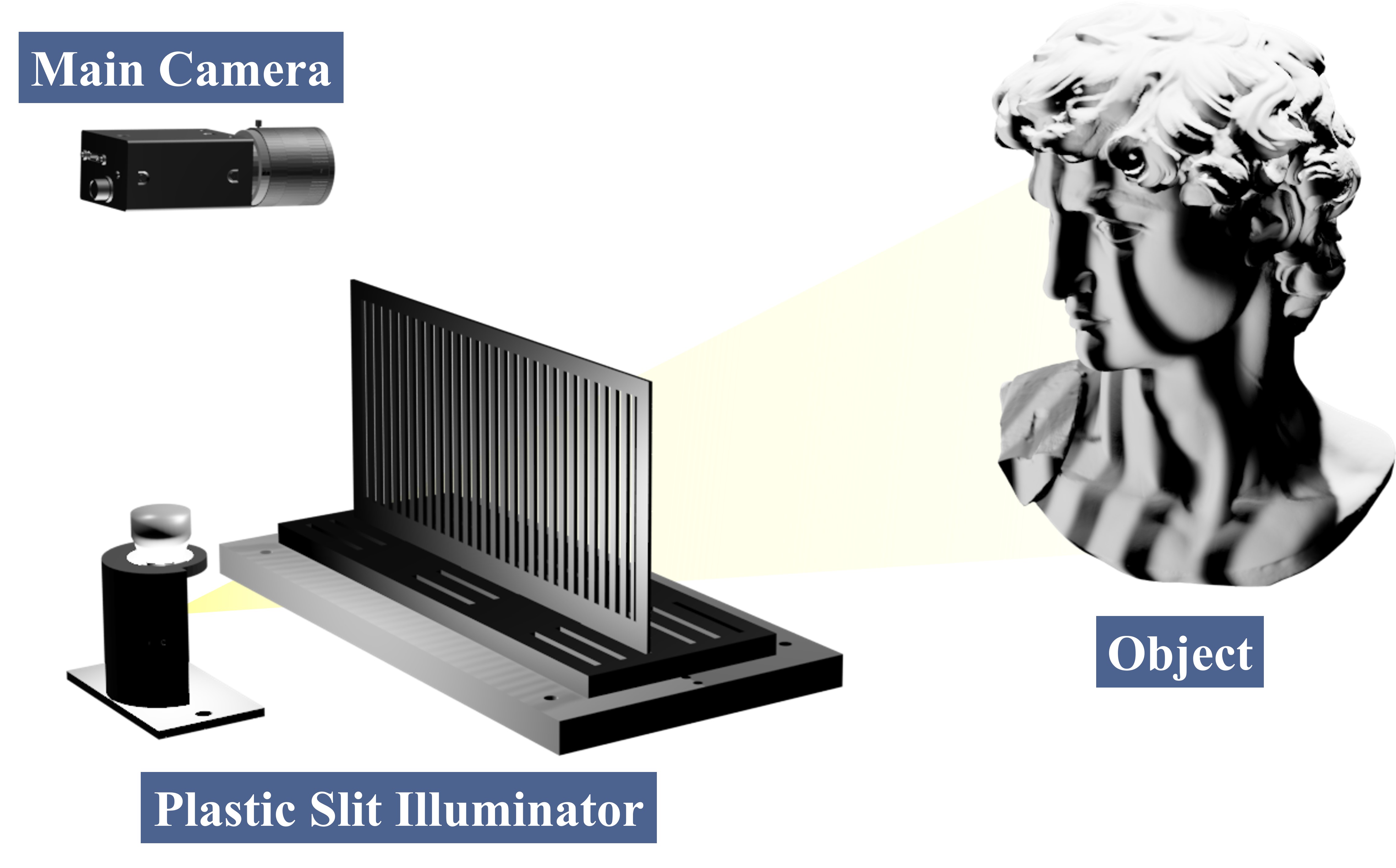}
 \caption{System schematic of reconstructing the 3-D geometry with the plastic slit illumination system after calibrating it with the proposed method. The 3-D geometry can be reconstructed only with a single camera and a plastic slit illumination system.}
 \label{fig:plasticStatueSchematic}
 \end{figure}

 \begin{figure}[!htb]
 \centering\includegraphics[width = 0.9\textwidth]{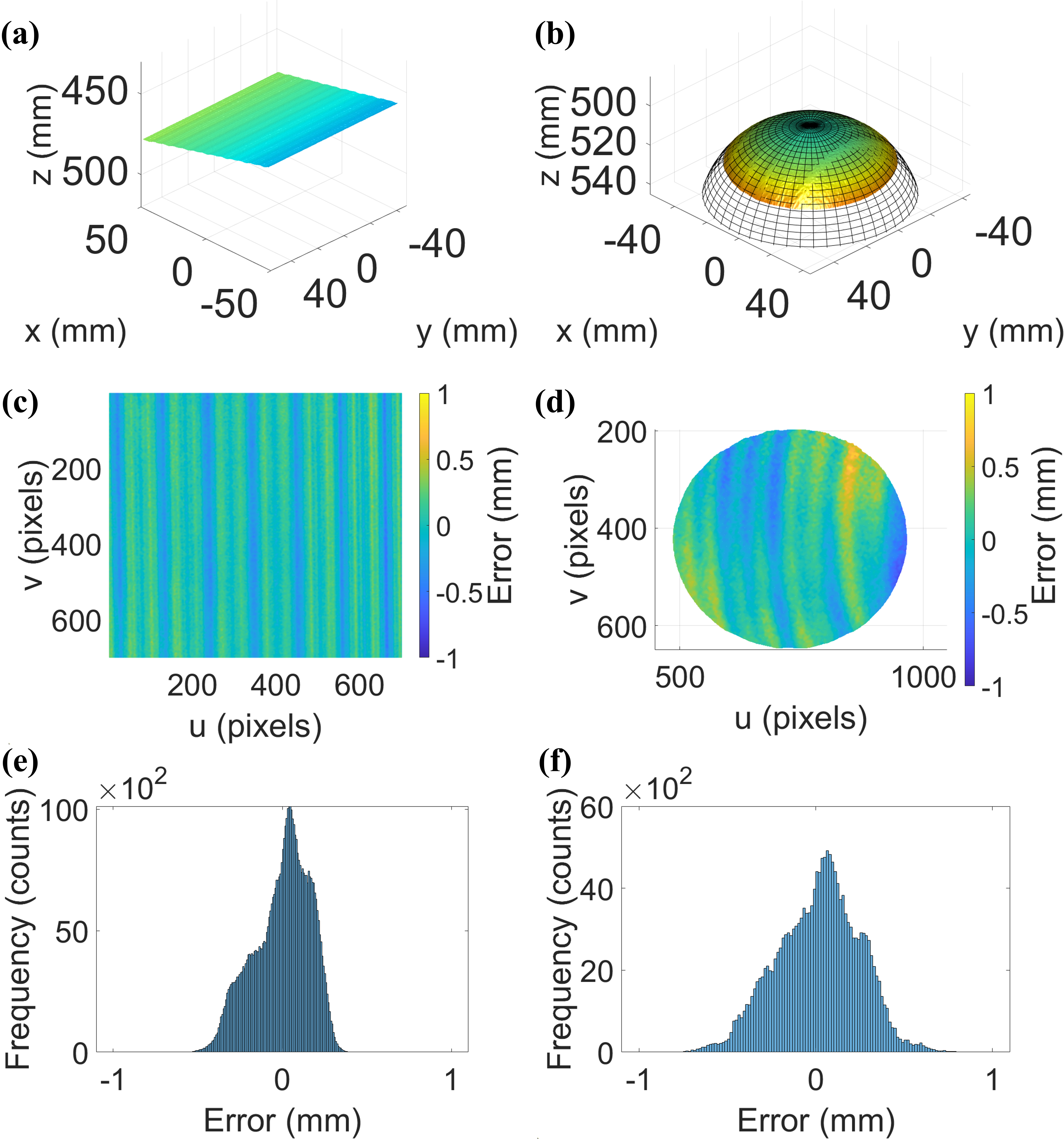}
 \caption{3-D reconstruction results using a single camera and a plastic slit illuminator. (a) 3-D reconstruction of the flat plane taken with the plastic slit illumination system calibrated by the proposed method. (b) 3-D reconstruction of a sphere with a radius of 50 mm taken with the plastic slit illumination system calibrated by the proposed method. (c) Offset error map from the ideal plane to the measurement made with the plastic slit illumination system calibrated by the proposed method. (d) Offset error map from the ideal sphere to the measurement taken with the plastic slit illumination system calibrated by the proposed method. (e) Histogram of the offset error map of reconstructed flat plane. (f) Histogram of the offset error map of reconstructed sphere.}
 \label{fig:offsetPLA}
 \end{figure}

 \begin{figure}[!htb]
 \centering\includegraphics[width = 0.9\textwidth]{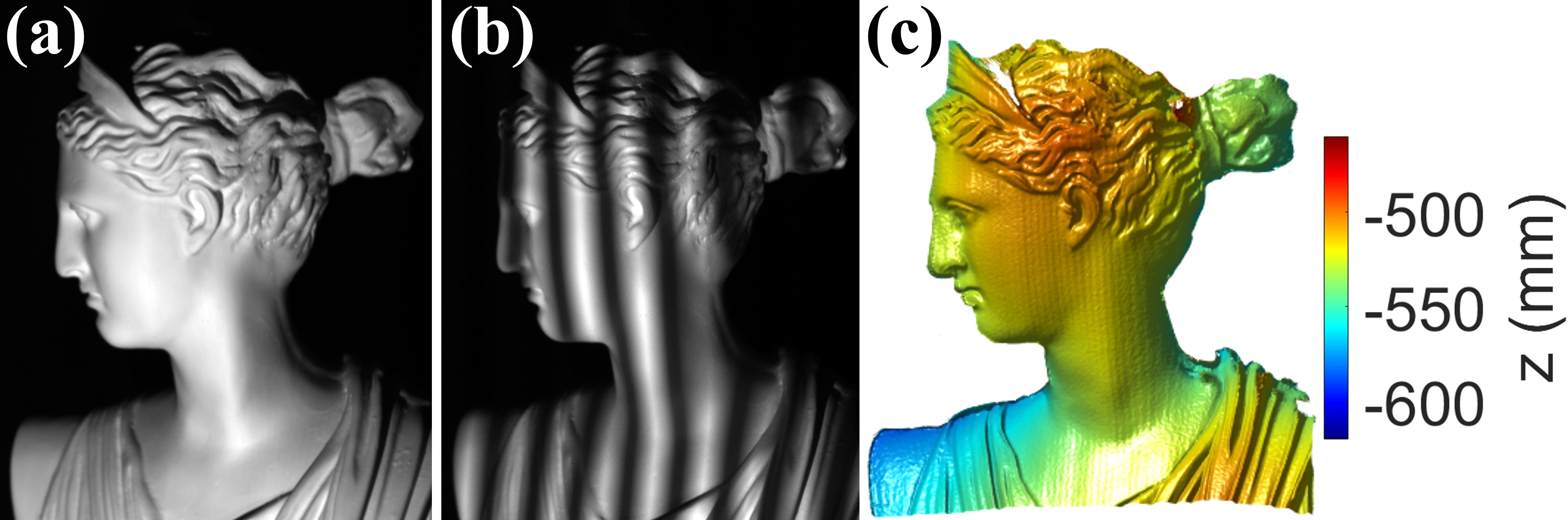}
 \caption{Reconstruction result of the statue taken with the plastic slit illumination system. (a) Texture image of the statue. (b) Fringe projected image of the statue. The fringe was generated by the plastic slits and the light source, in this case, a headlamp. (c) 3-D reconstruction of the statue taken with the plastic slit illumination system. The system was calibrated using the proposed method.}
 \label{fig:plasticStatue}
 \end{figure}

 To validate whether the proposed calibration method can also calibrate the system with a rough type illuminator, the digital light processing (DLP) projector in the previous experiment setup was replaced by a LED with a plastic slit. Fig.~\ref{fig:plasticSetup} shows the experimental setup of the system and Fig.~\ref{fig:plasticSlit} shows the plastic slits that were used in the experiment. When calibrating the system using the proposed method, the patterns were projected by placing the plastic slit in front of the headlamp. To project phase-shifted patterns, the slits were moved one by one and captured with cameras. The gray code patterns were also projected to unwrap the phase map using the plastic slits, which have been cut according to the shape of the gray code pattern. Fig.~\ref{fig:plasticPatterns} describes how the slits were shifted for pattern generations. To remove the phase discontinuities caused by machining error of the slits, the phase values of misaligned pixels were replaced with the median filtered phase values of the nearby pixels. In this experiment, 10 poses of the calibration board were taken with the plastic slit illumination system to calibrate the system. Fig.~\ref{fig:plasticFringes} shows the pattern-projected images of the calibration board and corresponding phase map. After the plastic slit illumination system is calibrated with the proposed method, the system can reconstruct a geometry with a simple configuration as shown in Fig.~\ref{fig:plasticStatueSchematic}. In order to evaluate the performance of the plastic slit illumination system calibrated with the proposed method, a mirror, sphere, and complex statue were taken with the system. Fig.~\ref{fig:offsetPLA}(a) and Fig.~\ref{fig:offsetPLA}(b) show the reconstruction result of a flat plane and a sphere repectively. Fig.~\ref{fig:offsetPLA}(c) and Fig.~\ref{fig:offsetPLA}(d) show the corresponding error map, which presents the offset from the measurement to the ideal geometry. The RMS error of the reconstruction result of the flat plane was 0.1647 mm. The measured radius of the sphere was 50.06 mm which has an error of 0.12 \% from the ground truth radius of 50 mm. The RMS error of a sphere reconstruction was 0.2521 mm. Fig.~\ref{fig:offsetPLA}(e) and Fig.~\ref{fig:offsetPLA}(f) show the corresponding histogram of each error map. Fig.~\ref{fig:plasticStatue} presents the result of reconstructing the complex statue with the plastic slit illumination system.

\section{\label{sec4}Conclusion}
In this paper, we introduced a novel calibration method for the FPP system, which skips the projector calibration owing to utilization of an auxiliary camera. It is later removed so that the system can remain a simple one-camera and one-projector configuration for reconstructing the geometry. The conventional FPP calibration method requires both horizontal and vertical direction fringes to be projected. The new method requires only one-directional fringe patterns, which halves the time for acquiring calibration data at each pose. In experiments, we calibrated the FPP system with only vertical patterns while ensuring the same level of accuracy in reconstructing the geometry. Moreover, unlike the conventional FPP calibration method, which is limited to DLP projectors, the proposed method can calibrate any fringe projection system even if it does not follow a pinhole model. In the second experiment, we proved that new method could calibrate an FPP system employing a rough type illuminator composed of a LED and plastic slits.


\section*{Acknowledgments}
The authors are thankful for the support received from Meta Reality Labs and Yonsei University (2023-22-0434).



 \bibliographystyle{elsarticle-num} 
 \bibliography{calib}





\end{document}